\documentclass[runningheads]{llncs}
\usepackage{graphicx}
\begin{document}
\title{Analyzing the spatial distribution of acute coronary syndrome cases using synthesized data on arterial hypertension prevalence \thanks{This research is financially supported by The Russian Science Foundation, Agreement \#19-11-00326. }}
\titlerunning{Analyzing the spatial distribution of acute coronary syndrome$\dots$}
%
\author{Vasiliy N. Leonenko \inst{1}\orcidID{0000-0001-7070-6584} }
\authorrunning{V. N. Leonenko}
%
\institute{ITMO University, 49 Kronverksky Pr., St. Petersburg, 197101, Russia
\email{vnleonenko@yandex.ru}}
\maketitle              
\begin{abstract}
In the current study, the authors demonstrate the method aimed at analyzing the distribution of acute coronary syndrome (ACS) cases in Saint Petersburg using the synthetic population approach and a statistical model for arterial hypertension prevalence. The cumulative number of emergency services calls in a separate geographical area (a grid cell of a map) associated with ACS is matched with the assessed number of dwellers and individuals with arterial hypertension, which makes it possible to find locations with excessive ACS incidence. The proposed method is implemented in Python programming language, the visualization results are shown using QGIS open software. Three categories of locations are proposed based on the analysis results. The demonstrated method might be applied for using the statistical assessments of   hidden health conditions in the population to categorize spatial distributions of their visible consequences. 

\end{abstract}
\section{Introduction}

Acute coronary syndrome (ACS) is a range of health conditions assosiated with a sudden reduced blood flow to the heart. This condition is treatable if diagnosed quickly, but since the fast diagnostics is not always possible, the death toll of ACS in the world population is dramatic \cite{jan2016catastrophic}. The modeling approach for forecasting the distribution of ACS cases would allow the healthcare specialists to be better prepared for the ACS cases, both in emergency services and in stationary healthcare facilities \cite{derevitskiy2017simulation}. The most simple forecast could be introduced by the application of the statistical analysis to the retrospective EMS calls data associated with acute heart conditions. However, if the corresponding time series dataset is not long, the accurate prediction is impossible without using additional data related to the possible prerequisites for acute coronary syndrome calls, such as health conditions that increase the risk of ACS.


One of the factors in the population which might be associated with acute coronary syndrome is arterial hypertension (or, shortly, AH) --- a medical condition associated with elevated blood pressure \cite{who_hyper}. Arterial hypertension is one of the main factors leading to atherogenesis and the development of vulnerable plaques whose instability or rupture are responsible for the development of acute coronary syndromes \cite{picariello2011impact}. Thus, we might assume that the neighborhood, which is populated predominantly by individuals with AH, might demonstrate higher vulnerability to ACS. Based on that assumption, it might be possible to use spatially explicit AH data as an additional predictor of prospective ACS cases. Unfortunately, the data on AH prevalence with the geographical matching are rarely found, and for Russian settings, they are virtually non--existent.

In this paper, we describe methods and algorithms to analyze the distribution of ACS--associated emergency medical service calls (shortly, EMS calls) using synthesized data on arterial hypertension prevalence. Using Saint Petersburg as a case study, we address the following question: may the synthesized AH data combined with EMS calls dataset provide additional information connected with ACS distribution in the population, compared to absolute data and relative data on EMS calls alone?

%

\section{Data}

\subsection{EMS calls} \label{ems_section}

\begin{figure}[htbp]
\centering
\includegraphics[width=1.1\textwidth, clip, keepaspectratio]{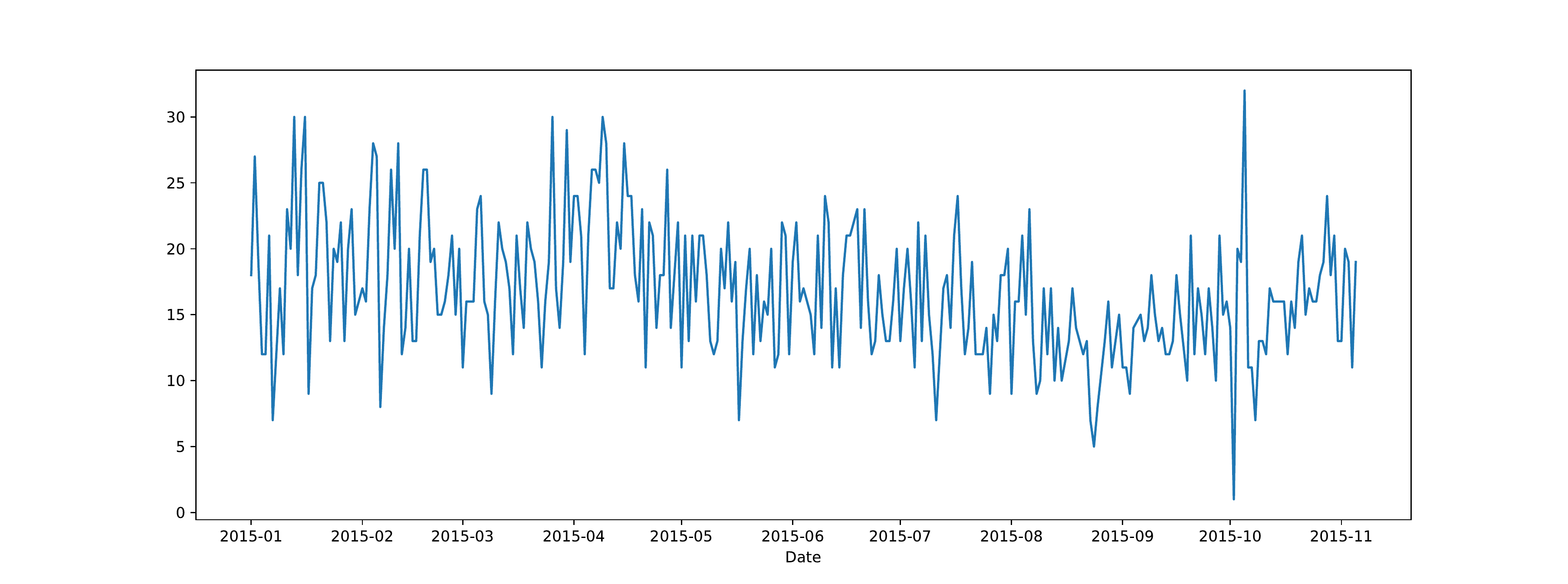}
\caption{The daily dynamics of emergency service calls connected with acute coronary syndrome (Jan -- Nov, 2015)}
\label{dynam_ems}       
\end{figure}

\begin{figure}[htbp]
\centering
\includegraphics[width=0.7\textwidth, clip, keepaspectratio]{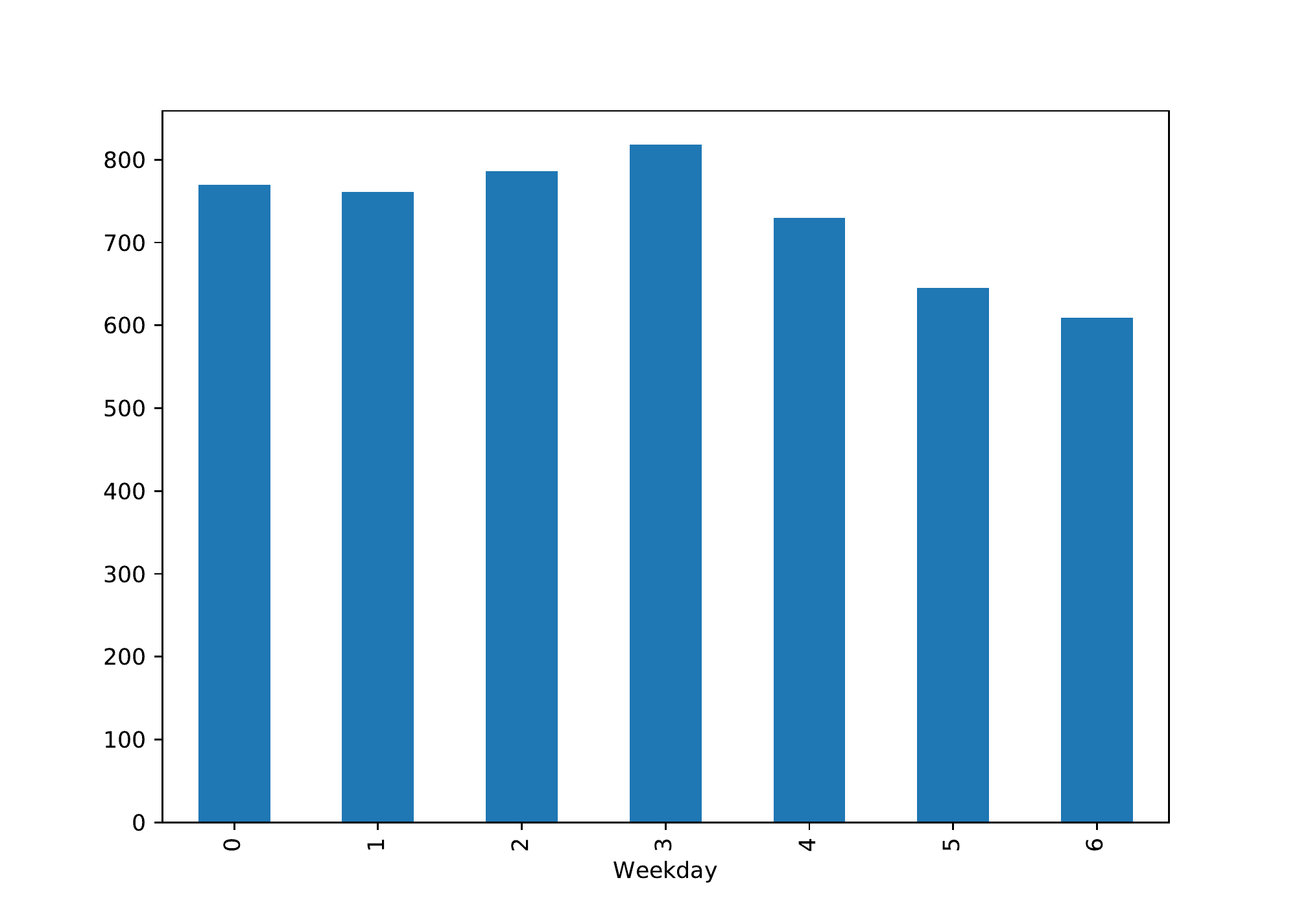}
\caption{The cumulative number of ACS emergency service calls in different weekdays.}
\label{ems_weekdays}       
\end{figure}

\begin{figure}[htbp]
\centering
\includegraphics[width=0.6\textwidth, clip, keepaspectratio]{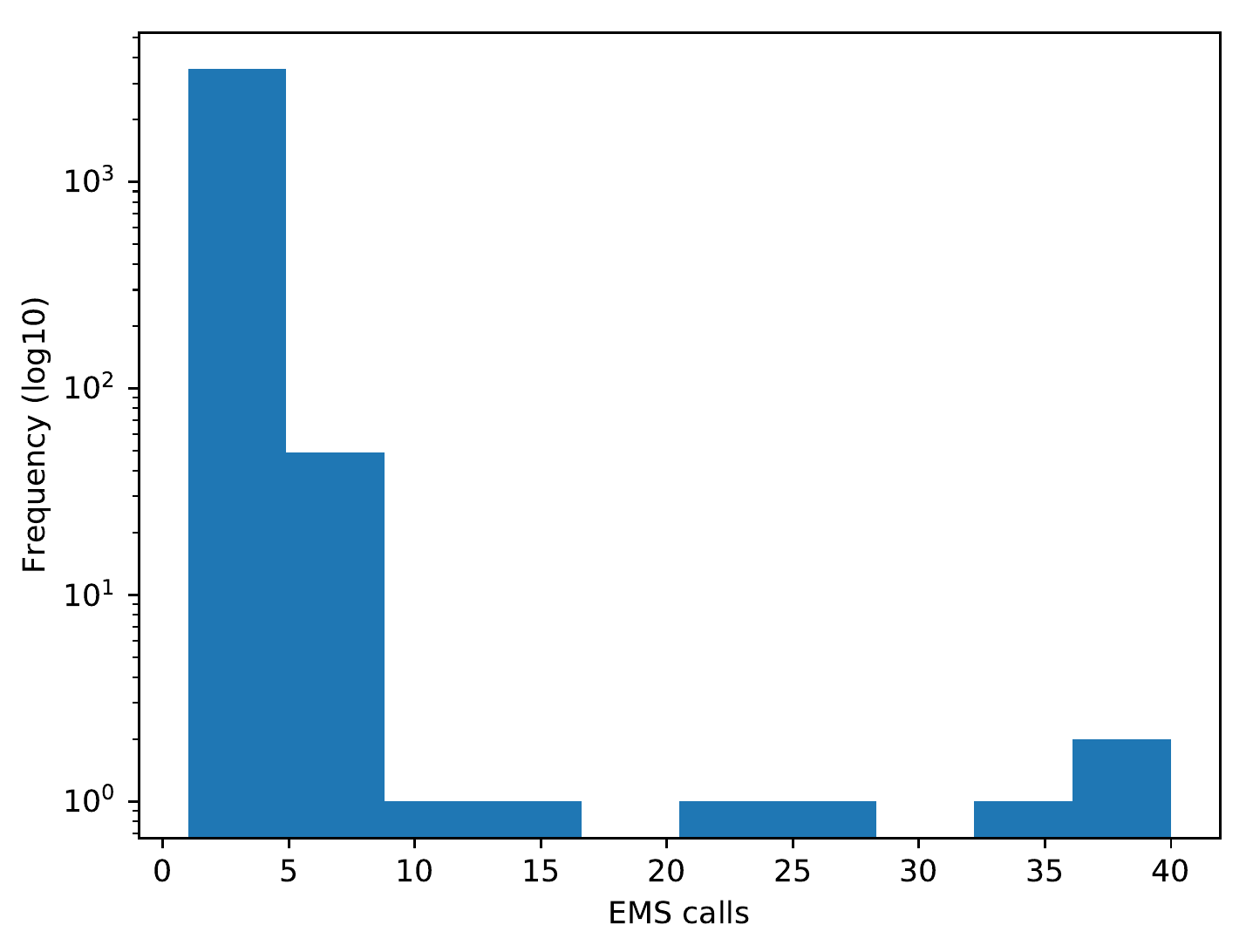}
\includegraphics[width=0.6\textwidth, clip, keepaspectratio]{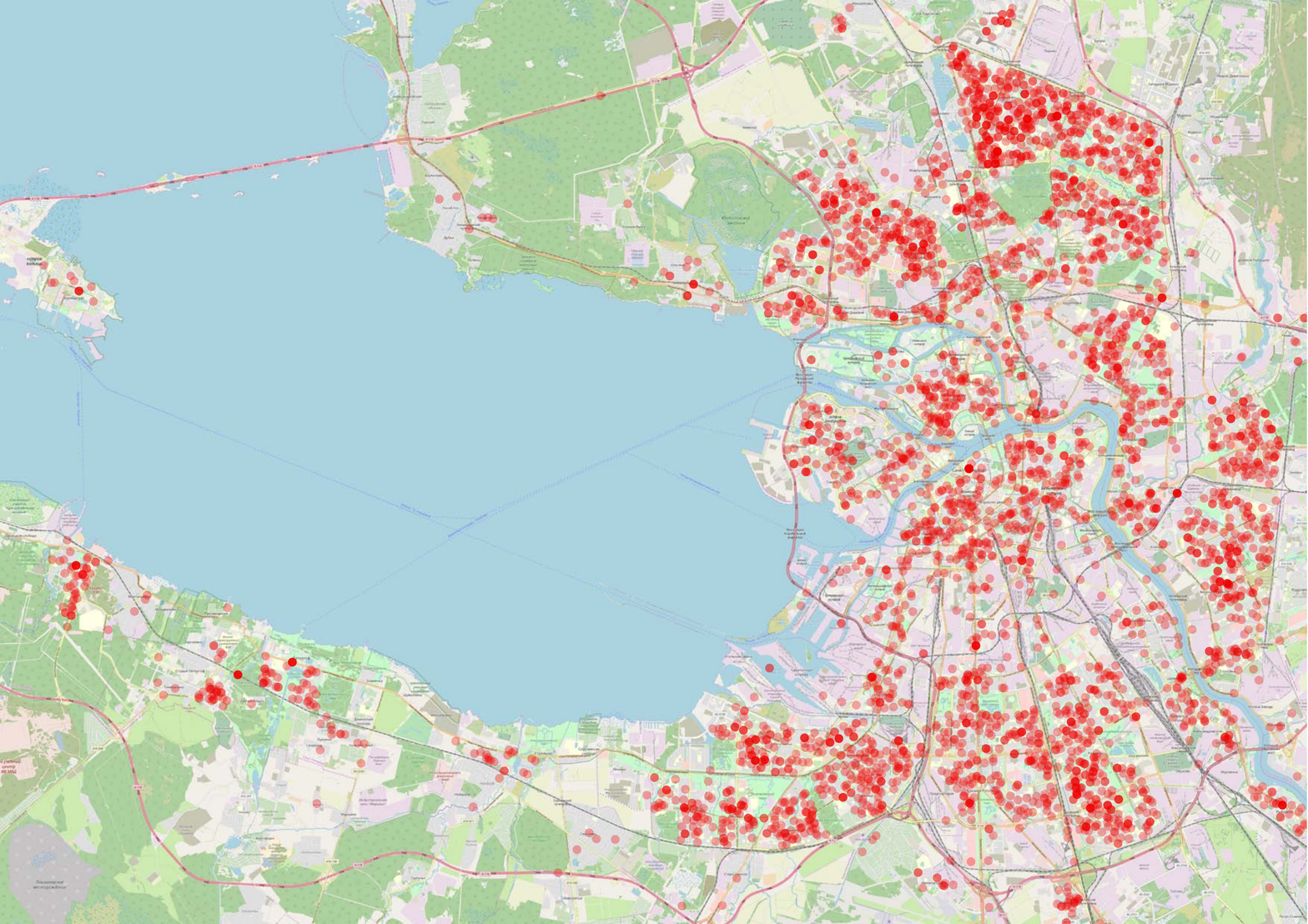}

\caption{The spatial distribution of ACS emergency service calls in Saint Petersburg}
\label{distr_ems}       
\end{figure}

The EMS data we used in the research contain 5125 EMS calls from January to November 2015 in Saint Petersburg connected with acute coronary syndrome  \cite{kovalchuk2018towards}. The back--of--the--envelope analysis of the time series corresponding to daily number of calls (Fig. \ref{dynam_ems}) and the weekly EMS calls distribution (Fig. \ref{ems_weekdays}) did not reveal any statistically significant patterns connected with distribution of calls over time, although it is clear from the data that the number of EMS calls has a decline in the weekends. Thus, there is no straightforward prediction method to forecast fluctuations of the cumulative number of daily EMS calls connected with ACS.

The spatial distribution of calls for the whole time period based on the addresses from the database is shown in Fig. \ref{distr_ems}. The histogram for cumulative distribution was built by calculating the total number of EMS calls in a given spatial cell with the size 250 m x 250 m, with empty cells (0 EMS calls) excluded from the distribution. It was established that the form of the histogram does not change significantly if the cell sizes vary (up to 2 km x 2 km). It can be seen that the predominant majority of the spatial cells had 1 to 5 EMS calls, and only for single cells this number exceeds 8. Based on general knowledge, we assumed that the increased concentration of the EMS calls within particular cells may be caused by one of the following reasons:

\begin{itemize}
\item The cell has higher population density compared to the other cells;
\item The cell has higher concentration of people with arterial hypertension, which might cause higher ACS probability;
\item The cell includes people more prone to acute coronary syndrome due to unknown reasons.
\end{itemize}

To distinguish these cases and thus to be able to perform a more meaningful analysis of EMS calls distribution, we assess the spatial distribution of city dwellers and people with high blood pressure using the synthetic population approach.

\subsection{Synthetic population}

A ``synthetic population'' is a synthesized, spatially explicit human agent database (essentially, a simulated census) representing the population of a city, region or country. By its cumulative characteristics, this database is equivalent to the real population, but its records does not correspond to real people. Statistical and mechanistic models built on top of the synthetic populations helped tackle a variety of research problems, including those connected with public health. In this study, we have used a synthetic population generated according to the standard of RTI International \cite{wheaton2009synthesized}. 

According to the standard of RTI International, the principal data for any given synthetic population is stored in four files: \texttt{people.txt} (each record contains id, age, gender, household id, workplace id, school id), \texttt{households.txt} (contains id and coordinates), \texttt{workplaces.txt} (contains id, coordinates and capacity of the workplaces), and \texttt{schools.txt} (contains id, coordinates, capacity). Our synthetic population is based on 2010 data from ``Edinaya sistema ucheta naseleniya Sankt Peterburga'' (``Unified population accounting system of Saint Petersburg'') \cite{unified_population_data}, which was checked for errors and complemented by the coordinates of the given locations. The schools records were based on the school list from the official web--site of the Government of Saint Petersburg \cite{spbgov}. The distribution of working places for adults and their coordinates were derived from the data obtained with the help of Yandex.Auditorii API \cite{auditorii}. The detailed description of the population generation can be found in \cite{leonenko2019spatial}.

\subsection{Assessing AH risk and individual AH status }

\begin{figure}[htbp]
\centering
\includegraphics[width=0.5\textwidth, clip, keepaspectratio]{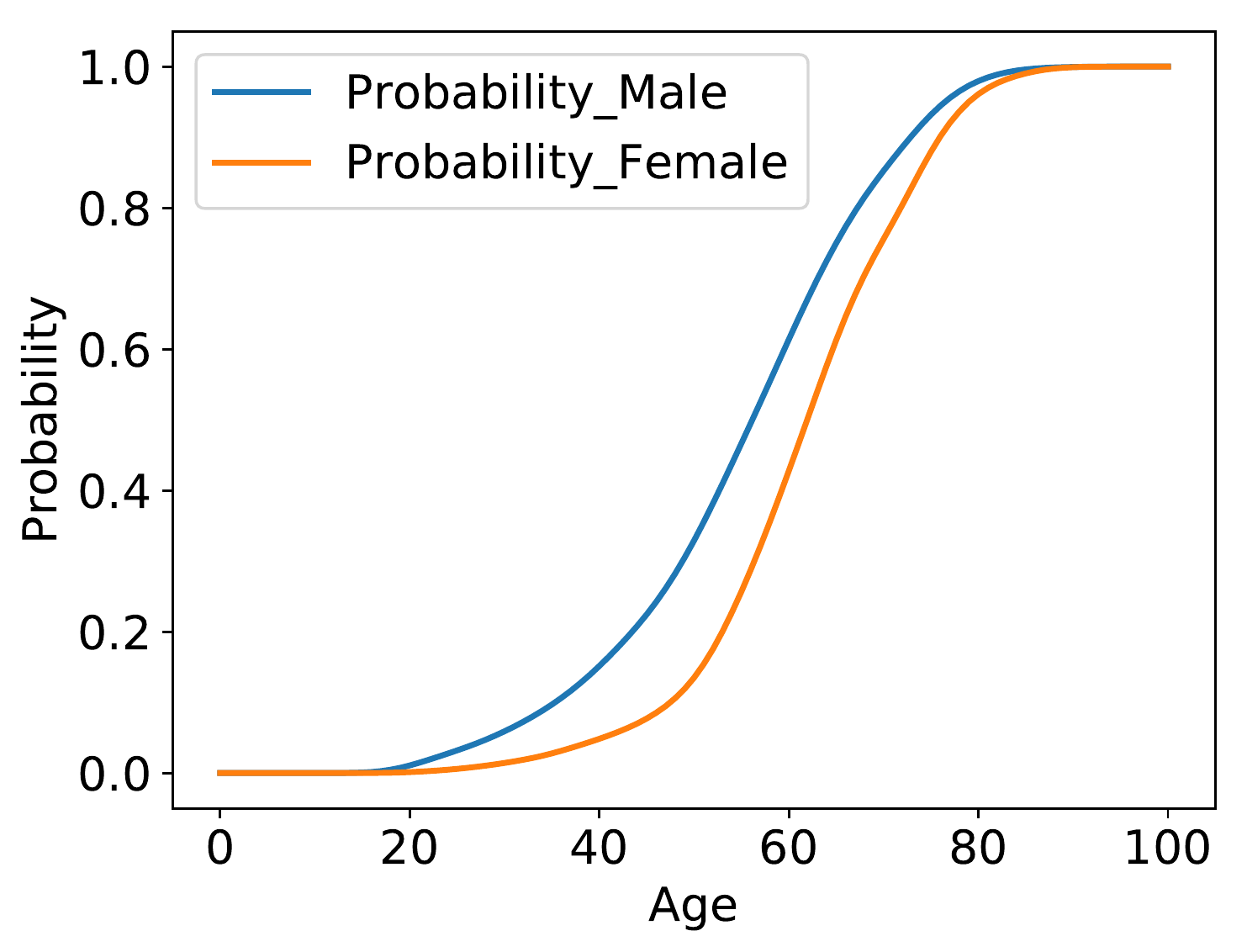}
\caption{The cumulative distribution function used to define the AH status of an individual, based on data from \cite{semakova2018}. }
\label{fig_cdf}       
\end{figure}

When the synthetic population is created, we assess the health conditions of individuals associated with arterial hypertension. There are two types of corresponding data that we generate and add to the individual records of the synthetic population: 

\begin{itemize} 
\item The AH risk (the probability of having arterial hypertension). Based on \cite{semakova2018}, we assumed that the mentioned probability depends on age and gender of an individual. The corresponding cumulative distribution function was found using the data of 4521 patients during 2010--2015 and is shown in Fig. \ref{fig_cdf}.

\item The actual AH status (positive or negative). The corresponding value (0 or 1) is generated by the Monte Carlo algorithm  according to the AH risk calculated in the previous step. The AH status might be used in simulation models which include demographic processes and population-wide simulation of the onset and development of AH. 
\end{itemize}

\begin{figure}[htbp]
\centering
\includegraphics[width=0.8\textwidth, clip, keepaspectratio]{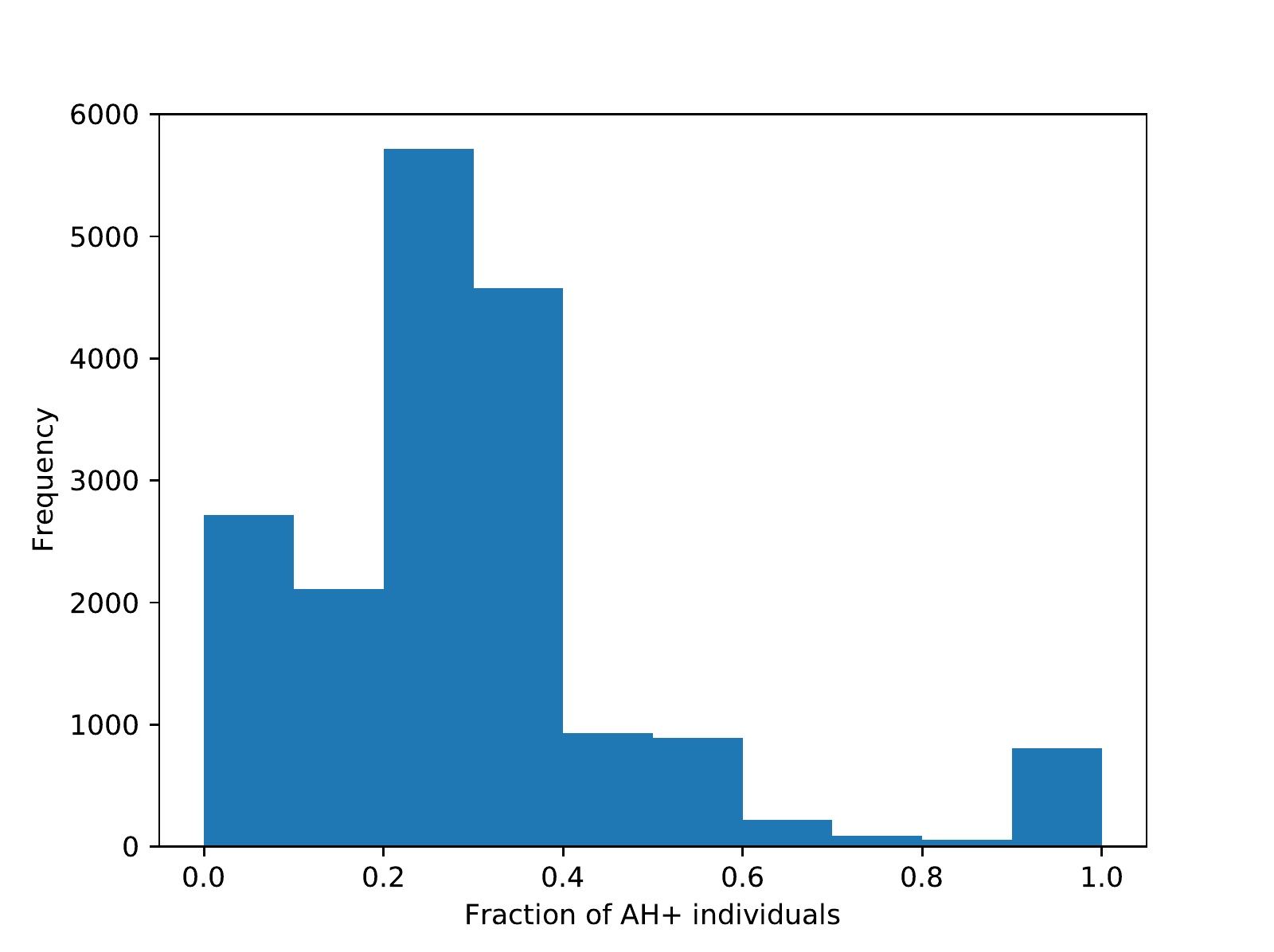}
\includegraphics[width=0.8\textwidth, clip, keepaspectratio]{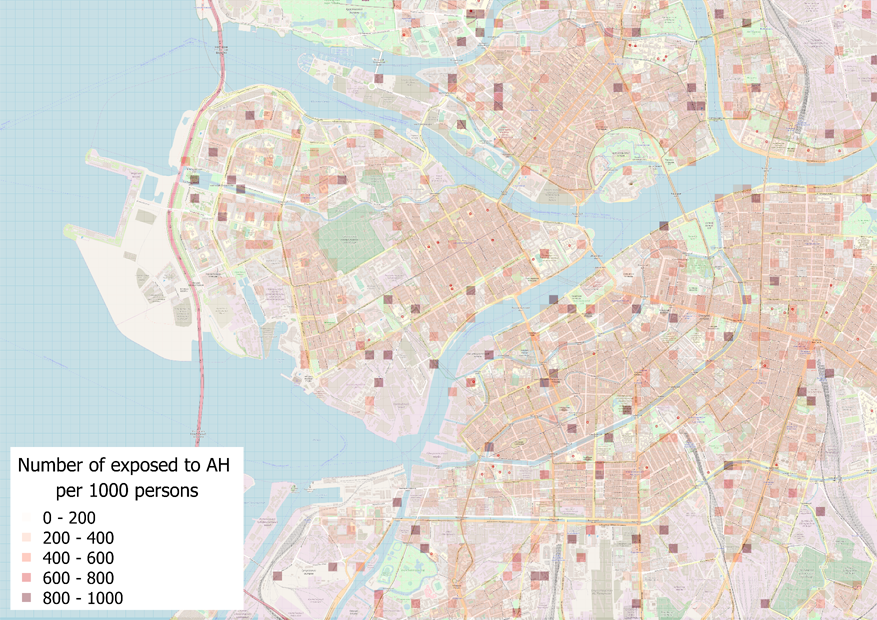}
\caption{The aggregated and geospatial distributions of AH+ individuals in Saint Petersburg}
\label{fig_squares}       
\end{figure}

The proportion of the synthetic population affected by arterial hypertension is found to be 26.6 \% which roughly correlates with the AH prevalence data in the USA according to American Heart Association Statistical Fact Sheet 2013 Update (1 out of every 3) and is lower than the estimate for Russia of the main cardiologist of Ministry of Health of Russian Federation (43\%). The cumulative and spatial distributions of AH+ individuals in Saint Petersburg  are shown in Fig. \ref{fig_squares}. It can be seen, that the age and gender heterogeneity in the population is enough to create uneven distribution of individuals exposed to arterial hypertension.

Further in the paper we match the number of AH+ dwellers of every cell with the number of EMS calls within this same cell and propose an indicator to analyze the relation between them.

\subsection{Calculating the indicators related to EMS calls.} We convert the coordinates of EMS calls location from degrees to meters using Mercator projection. After this we form a grid with a fixed cell size (250m $\times$ 250m), defining its bounds with maximum and minimum coordinates of the synthetic individuals. Finally, using the EMS calls dataset, we calculate the overall number of EMS calls for each cell of the grid. In the same way, we calculate the overall number of dwellers and AH+ individuals for the cells. This algorithm was implemented as a scripts collection written in Python 3.6 with the libraries \texttt{numpy}, \texttt{matplotlib}, and \texttt{pandas}. The output of the algorithm is a txt-file with the coordinates of the cells and the cell statistics (overall number of individuals, number of AH+ individuals, overall number of EMS calls). 

In order to understand the relationship between the numbers of AH+ users and the number of EMS calls, we follow our earlier research \cite{bates2018using}, where the ratio $r_1$ between the overdose--related EMS calls and the assessed number of opioid drug users was studied. In this paper, we compare $r_1$ with the alternative indicator $r_2$ which uses the number of people in overall in the cell under study instead of the assessed quantity of AH+. The formulas to calculate the following ratios are the following:
$$
r_1=\frac{n_{ems}+1}{n_{ah}+1} {\rm \hspace*{0.5cm} and \hspace*{0.5cm}} r_2=\frac{n_{ems}+1}{n_{p}+1}
$$
where $n_{ems}$ is the number of registered EMS calls in a cell, $n_{ah}$ is the model--predicted number of AH+ users in a cell, and $n_{p}$ is the number of persons who dwell in a cell based on the synthetic population. These quantities represent ratios of calls per AH+ individual and calls per dweller, respectively. By adding ones to the numerator and denominator we are able to avoid a divide by zero error, and although it provides a small skew in the data, its consistent application across all cells leaves the results and their interpretations unhindered.  We use the ratio $r_1$ to understand which cells have large differences in the orders of magnitude compared to other cells. The ratio $r_2$ is introduced to compare its distribution with $r_1$ and thus decide whether the statistical model for AH+ probability assessment helps more accurately detect the anomalies connected with EMS calls distribution.
 
\section{Results}

\subsection{Cumulative distribution}
\begin{figure}[htbp]
\centering
\includegraphics[width=0.47\textwidth, clip, keepaspectratio]{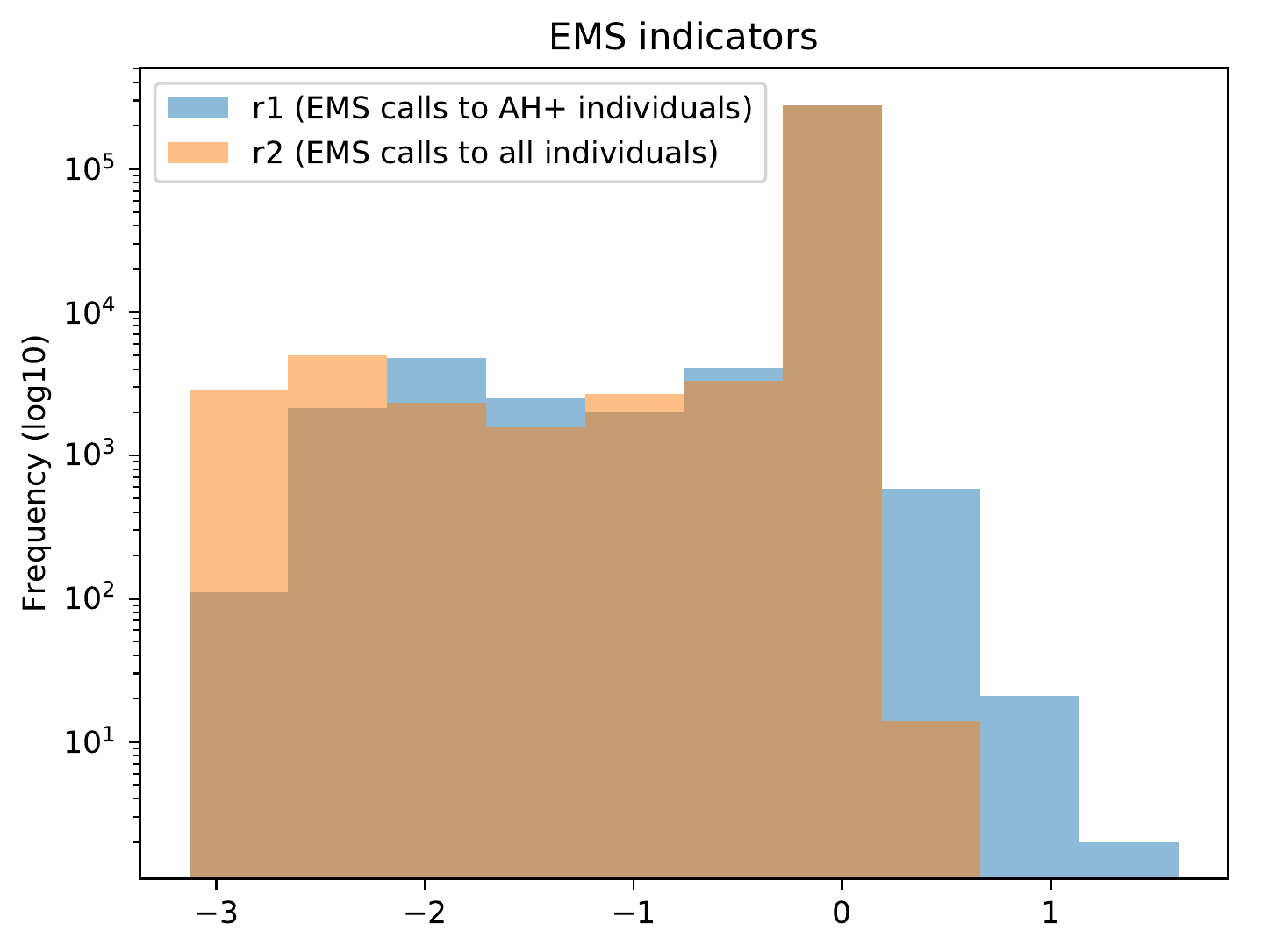} 
\includegraphics[width=0.47\textwidth, clip, keepaspectratio]{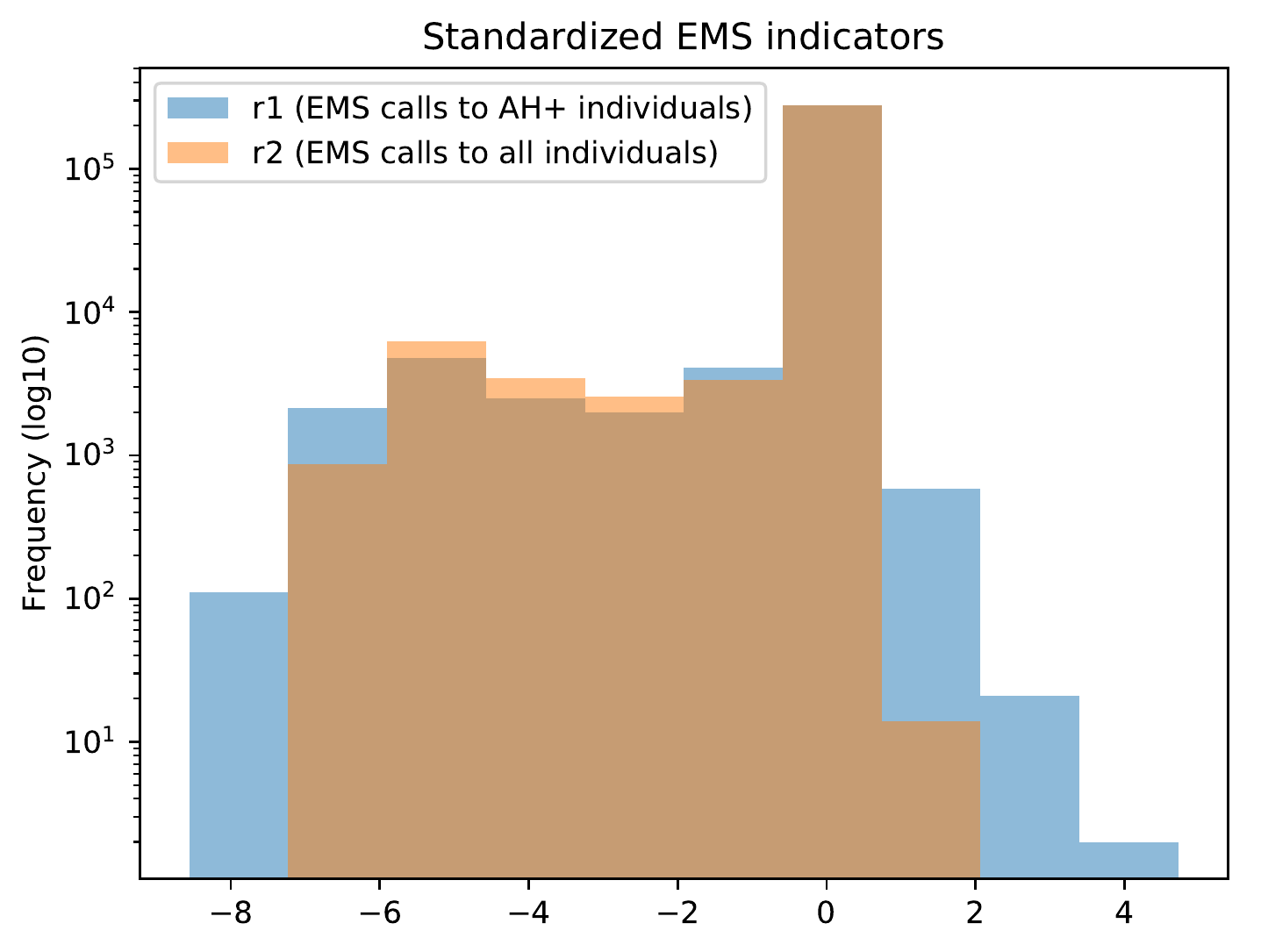}
\caption{The distributions of $r_1$ and $r_2$ (original and standardized). }
\label{fig_r1_r2}       
\end{figure}

%


In Fig. \ref{fig_r1_r2}, the aggregated distributions of the $r_1$ and $r_2$ values for our data are shown. On the left graph, the distributions are given in their original form, and in the right one the standardized distributions are demonstrated, i.e. with means equal to 0 and standard deviations equal to 1. Although the shape of the histograms is similar, the difference between the corresponding distributions is statistically significant, which is supported by the results of Chi--square test performed for the standardized samples. The crucial difference is in the histogram tails, i.e. in the extreme values of the indicators, which, as it will be shown further in the paper, is also accompanied by their different spatial distribution. 

\subsection{Spatial distribution}

\begin{figure}[htbp]
\centering
\includegraphics[width=1.0\textwidth, clip, keepaspectratio]{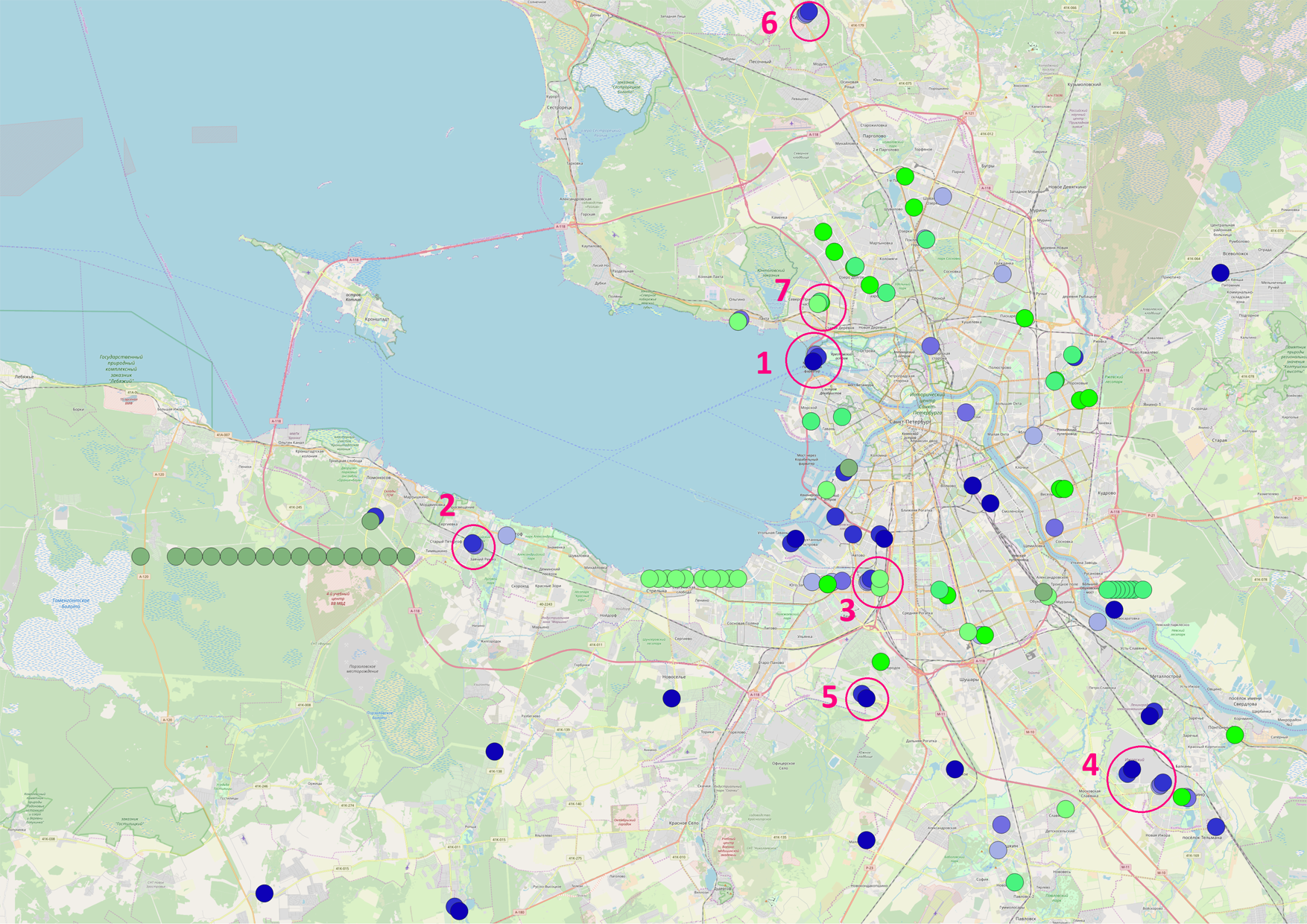} 
\caption{Points of high $r_1$ and $r_2$}
\label{fig_top20}       
\end{figure}

\begin{figure}[htbp]
\centering
\includegraphics[width=0.9\textwidth, clip, keepaspectratio]{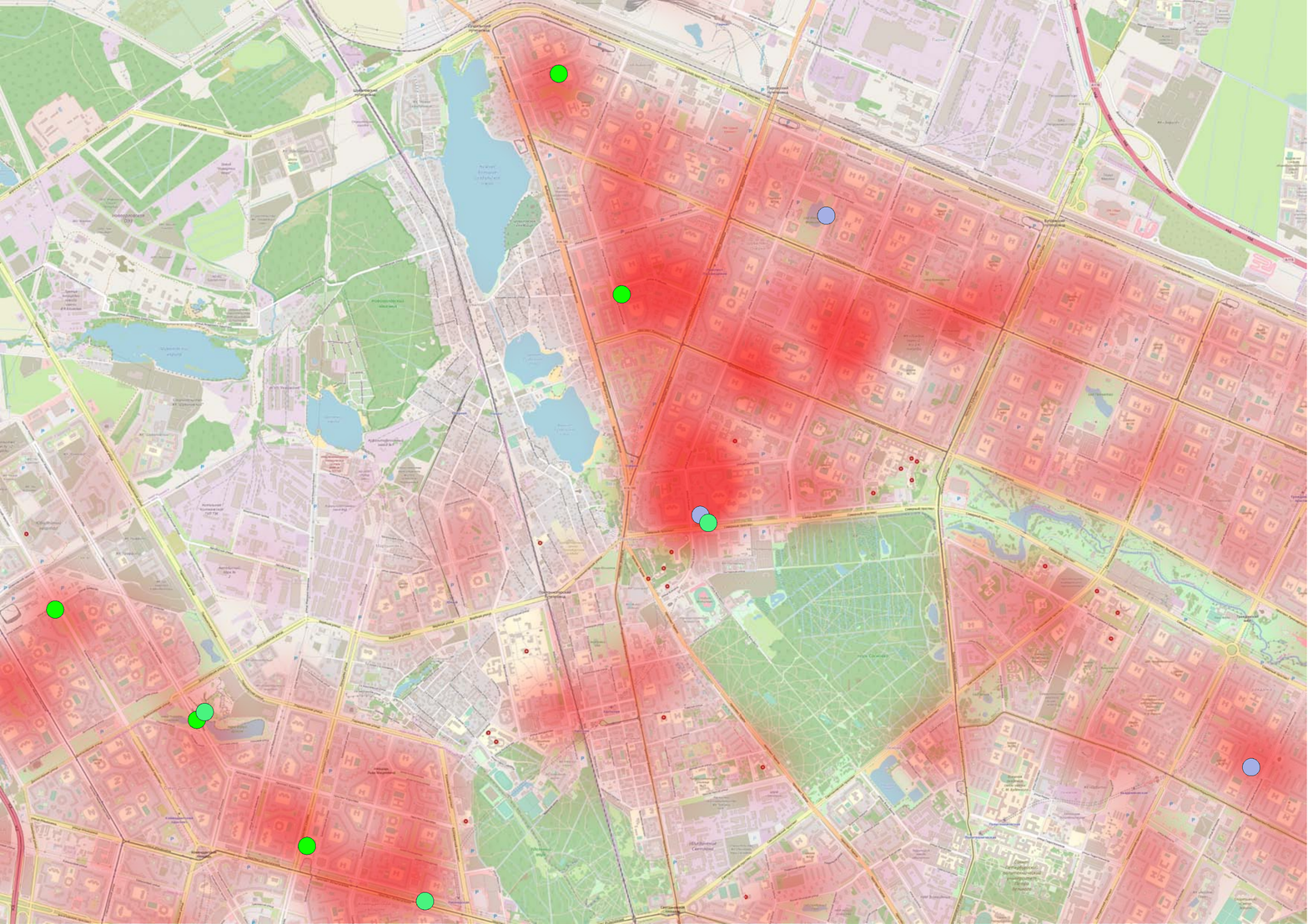} 
\caption{Heatmap of EMS calls matched against high $r_1$ and $r_2$ locations}
\label{fig_ems_heatmap}       
\end{figure}

In Fig. \ref{fig_top20}, a distribution of 20 cells with the highest values of $r_1$ and $r_2$ is shown (shades of blue and shades of green correspondingly). The lighter shades corresponds to the bigger cell side lengths (250, 500, 1000 and 2000 meters).

The results demonstrate that the locations of high $r_1$ values change less with the change of cell side length, compared to $r_2$ (it is represented in the map as several points with different shades of blue situated one near another). Also it is notable that the high $r_2$ values were found in lined up adjacent cells (see left and right edges of the map). This peculiarity of $r_2$ distribution requires further investigation, because it hampers the meaningful usage of the indicator.

The locations marked with three blue points represent concentration of high EMS calls in the isolated neighborhood with few assessed number of AH+ individuals. Most of these locations happen to be near the places connected with tourism and entertainment (1 -- Gazprom Arena football stadium, 2 -- Peterhof historical park) or industrial facilities (3 -- bus park, trolleybus park, train depot; 4 -- Izhora factory, Kolpino bus park). Location 5 corresponds to Pulkovo airport, a major transport hub (it is marked by only two blue points though). Location 6 is the one which cannot be easily connected with excessive EMS calls --- it is situated in a small suburb with plenty of housing. The possible interpretation of why it was marked is the discrepancy between the actual number of dwellers for 2015 (a year for EMS calls data) compared to the 2010 information (a year for populational data). This zone was a rapidly developing construction site which caused the fast increase in number of dwellers. Location 7 is also an expectational one -- it is the only one which is marked by three green points (high $r_2$). Additionally, this zone was not marked by high $r_1$, although it is easily interpreted as yet another industrial district (Lenpoligraphmash printing factory). Increasing the number of points in a distribution to 100 does not change significantly the results: still blue points mark isolated areas with meaningful interpretation (except Lenpoligraphmash at location 7 which is still solely marked by green).

Whereas the exceptional values of $r_1$ indicate isolated non--residential areas (industrial objects and places of mass concentration of people) which might be connected with the increased risk of ACS  and thus require attention from healthcare services, the extreme values of $r_2$ indicator might come in handy in the situation when we need to assess the excess of EMS calls in the densely populated residential areas. In Fig. \ref{fig_ems_heatmap}, where $r_1$ and $r_2$ values are plotted against a heatmap of EMS call numbers, we see that there are two types of peak concentrations of EMS calls (bright red color). Ones are not marked with green dots (the $r_2$ values are not high) and thus might be explained by high concentration of dwellers in general. Others, marked with green dots, show the locations with high number of EMS calls relative to population. In case when there is no corresponding high $r_1$ value, these spots might correspond to the category of neighborhoods with ACS risk factors not associated with arterial hypertension (to be more precise, not associated with the old age of dwellers, since it is the main parameter of the statistical model for AH prevalence used in this study).

\section{Discussion}
In this paper, we have demonstrated a statistical approach which with uses synthetic populations and statistical models of arterial hypertension prevalence to distinguish several cases of ACS--associated EMS call concentration in the urban areas:

\begin{itemize}
\item High $r_1$ values for any corresponding number of EMS calls (Fig. \ref{fig_top20}) might indicate areas where acute coronary syndrome cases happen despite the low AH+ population density and thus require attention from the healthcare organs.

\item Average to low $r_2$ values for high number of EMS calls (Fig. \ref{fig_ems_heatmap}, red spots without green points) correspond to areas with high population density.  

\item High $r_2$ values and low $r_1$ values for high number of EMS calls (Fig. \ref{fig_ems_heatmap}, red spots with green points) might indicate areas where the excessive number of ACS cases cannot be explained neither by the high population density, nor by AH prevalence, thus they might indicate neighborhoods with unknown negative factors. 

\end{itemize}

It is worth noting that due to the properties of our EMS dataset (see Section \ref{ems_section} and Fig. \ref{distr_ems}) most of the locations with extremely high $r_1$ and $r_2$ correspond to the number of EMS calls in a grid cell equal to 1. Ascribing EMS calls to one or another property of the area based on such a small number of observations is definitely premature, and thus our interpretations given earlier in the text should be continuously tested using the new data on EMS calls. Despite the fact that we cannot draw any definite and final conclusions, in the author's opinion, the study successfully introduces the application of the concept of using synthesized data related to mostly unobserved health conditions in the population (arterial hypertension) to categorize spatial distribution of their visible consequences that require immediate medical treatment (acute coronary syndrome). As it was demonstrated by the authors before \cite{bates2018using}, the same approach can be successfully used in case of opioid drug usage, and we expect to broaden the scope of its application by applying it in another domains.

As to the current research, we plan the following directions of its further development:

\begin{itemize}

\item Currently, the time periods of the EMS calls information and synthetic population data does not match which might cause the bias in the estimated values of the indicators. We plan to actualize both datasets and to establish whether the results which are demonstrated in this study will be reproduced.

\item The enhanced statistical model for AH is required to make the calculation of the number of AH+ individuals more accurate.

\item The values of $r_1$ are almost the same for the cases of (a) 1 EMS call in presence of 0 AH+ individuals, and (b) $2n$ calls in presence of $n$ AH+ individuals, so those cases cannot be distinguished by using indicators such as $r_1$, although they are essentially different. We want to explore the possibility of using a yet another indicator which will take into account the absolute number of dwellers in the neighborhood and will have a meaningful interpretation.

\item We have access to a number of health records of the people hospitalized with ACS in a human--readable format, which contains information about their AH status. Using natural language processing tools, we plan to obtain a digital version of this dataset and consequently to  assess numerically the connection between AH and ACS cases in Saint Petersburg. This result will help reduce uncertainty in the results of the current study connected with analyzing the distribution of $r_1$.

\end{itemize}

%
%
%
\bibliographystyle{splncs04}
\bibliography{leonenko}

\end{document}